\documentclass[preprint,aps,floats,nofootinbib,amssymb]{revtex4}
\pdfoutput=1 
\usepackage[T1]{fontenc} 
\usepackage{hyperref}
\usepackage[utf8]{inputenc}
\usepackage{amssymb}
\usepackage{latexsym}
\usepackage{graphics}
\usepackage{graphicx}
\usepackage{feynmf}
\usepackage{color}
\usepackage{epsf,epsfig}
\usepackage{amsmath}
\usepackage{hyperref}
\usepackage{subfigure}
\usepackage{mathrsfs}

\newcommand{\be}{\begin{equation*}}
\newcommand{\ee}{\end{equation*}}
\newcommand{\ba}{\begin{eqnarray*}}
\newcommand{\ea}{\end{eqnarray*}}

\newcommand{\bw}{\begin{widetext}}
\newcommand{\ew}{\end{widetext}}

\begin{document}         
\title{\vspace*{1.in}\large
Impact of $H_0$ on gravitational wave propagation:\\ prospects for the LISA mission}
\vspace*{0.5cm}

\author{Nil Blanco\footnote{nilblanco2003@gmail.com}  and Dom\`enec Espriu\footnote{espriu@icc.ub.edu}}
\affiliation{
Departament de F\'isica Qu\`antica i Astrof\'isica and
Institut de Ci\`encies del Cosmos (ICCUB), \\
Universitat de Barcelona, 
Mart\'i Franqu\`es 1, 08028 Barcelona, Spain}

\begin{abstract}
  It was previously shown that $H_0$ influences gravitational wave propagation beyond simple redshift, by modifying the effective wavenumber.
  While earlier studies focused on Pulsar Timing Arrays, we analyze the observability of this effect with LISA. Modeling the LISA arm geometry,
  we derive the timing residual and identify a clear angular enhancement at frequencies above 100 mHz. Using the stationary phase
  approximation and numerical methods, we show that the optimal incidence angle depends on $Z_A H_0$, providing a direct, redshift-independent
  measurement of the Hubble parameter. We estimate that LISA could determine $H_0$ to within a few percent using this method, independent
  of standard siren approaches.
\end{abstract}

\maketitle


\section{Introduction}

The detection of gravitational waves (GW) by LIGO and Virgo opened a new era in astrophysics,
allowing us to probe the universe in ways previously impossible.
While ground-based detectors operate at high frequencies, the future Laser Interferometer
Space Antenna (LISA) will predominantly operate in the milli-Hertz band\cite{LISAreview}, sensitive to
supermassive black hole mergers and other cosmological sources.

Earth-based interferometers have arms that are extremely short compared to the typical wavelength
of the wave they observe. For instance, in LIGO we should compare a 4 kilometer arm length with a
typical wavelength of about 600 km from a binary neutron star inspiral  merger that radiates with a frequency of 500 Hz.
In this sense, Earth-based interferometers should be called ``local''.

This implies that LIGO-VIRGO detect a variation in the optical path that exclusively
depends on the frequency. In other words, if we think of the optical path as being divided into small line
elements, all such elements periodically contract and expand in a similar manner and it is the cumulative effect
on the phase that is finally measured. 

The presence of the cosmological expansion
induces corrections in the equations of motion for metric perturbations. These corrections manifest {\em both}
in a redshift of
the frequency and a modification of the effective wave number. The analysis of the linearized perturbations
is simplest on a de Sitter background metric,
\begin{equation}
G_{\mu\nu}(\tilde{g}+h) = G_{\mu\nu}(\tilde{g}) + 
\frac{\delta G_{\mu\nu}}{\delta g_{\alpha \beta}}\, \biggl\rvert_{\tilde{g}}\ h_{\alpha\beta} + \ .\ .\ .\ 
= \Lambda \tilde{g}_{\mu \nu} + \Lambda h_{\mu \nu} ,
\end{equation}
($G_{\mu\nu}$ is the Einstein tensor) that translate in Friedmann-Lemaitre-Robertson-Walker (FLRW) comoving coordinates to the following wave equation
\begin{equation}\label{linearized}
\Box_{\text{FLRW}} \, h_{ij} \equiv \left[\partial^2_T - \sqrt{\frac{\Lambda}{3}} \, \partial_T - 
\frac{1}{a^2} \left(\partial^2_R + \frac{2}{R} \, \partial_R \right) - 2 \, \frac{\Lambda}{3} \right] h_{i j}= 0,
\end{equation}
which can be solved perturbatively in $\Lambda$. Here $a(T)$ is the cosmological factor and $R$ the comoving
distance.

A more detailed analysis proves that at leading order in the cosmic energy densities $\sqrt{\Lambda/3}$ can be replaced by
the Hubble parameter $H_0$ because all the cosmological
components  enter on an equal footing at this order\cite{espriualfaro}.
After choosing the transverse-traceless gauge, the solution will be a combination of trigonometric functions with
different frequencies. For simplicity let us just consider one of such waves
\begin{equation}\label{leading}
h_{ij}^{FLRW}(T,R)=\frac{\epsilon_{ij}}{R}\left(1+H_0T\right)\cos(\omega_{eff}T-k_{eff}R)
\end{equation}
Here \be
w_{eff}= w(1-RH_0)\qquad  k_{eff}=w(1- RH_0/2)
\ee
We note that $w_{eff}$ is just the usual redshifted frequency ($\simeq w/(1+z)$) when $z$ is treated perturbatively.
At the next order\cite{espriualfaro}, the contribution from the cosmological constant, matter and radiation densities appear
with different factors and the solution cannot be only expressed in terms of $H_0$. It is important to understand
that if one writes $k_{eff}= \omega_{eff}$, the resulting $h_{ij}$ is not a solution of the wave equation. 

As already said, GW have rather long wavelengths and $k_{eff}$ will produce a modulation of the signal along the arms of the
detectors provided that these are long enough. For instance, the merger of very massive black holes, with frequencies ranging
from $10^{-6}$ to $10^{-9}$ Hz, produce wavelengths of the order of light-years. 

\begin{figure}
\centering
\includegraphics[scale=0.85]{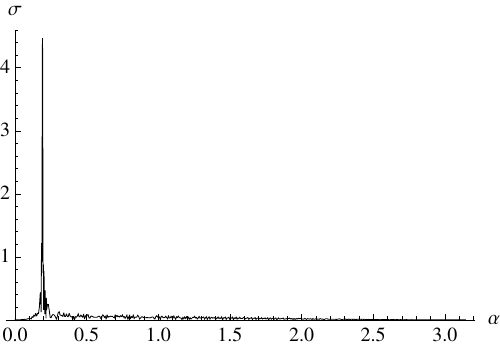}
\includegraphics[scale=0.6]{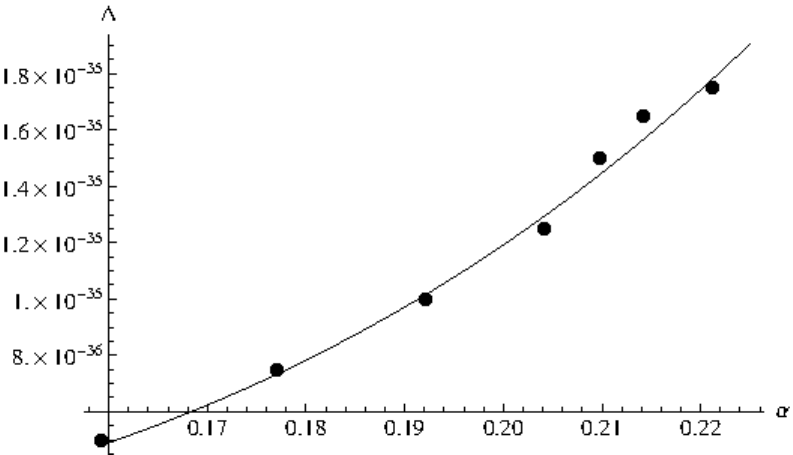}
\caption{PTA results. Left: the statistical significance of the enhancement $\sigma(\alpha)$ for $\Lambda=10^{-35}s^{-2}$.
  Right:$H_0(\alpha)$ obtained numerically
  from the positions of the peaks in the $\sigma(\alpha)$ plots for different values of $H_0$ (dots) and obtained analytically from an
  approximation of the Fresnel functions involved in the timing residual (continuous line).  }
\label{fig:pta}
\end{figure}

Previous works analyzed this effect\cite{espriupuigdomenech,espriualfaro,espriu2021,er2} in the context of Pulsar
Timing Arrays (PTA), where the baseline $L$, the Earth-pulsar distance,
is of the order of kiloparsecs (hence longer than the characteristic wavelength of very massive black hole mergers) and
it was found that the time residual had a marked enhancement for a specific value of the angle subtended by the source and the
pulsar-Earth arm. This enhancement depends only on the product of the comoving distance and the Hubble parameter.
In Figure \ref{fig:pta} we show an example of this enhancement and the dependence of the angle of the enhancement on the redshift
The enhancement is visible both in snapshots of the signal, for a fixed time, and also -much enhanced- after integration of the
signal over a period of time (years in the case of the PTA). The analytical  determination of $H_0$ from $\alpha$ will be discussed later
as it is exactly the same for PTA and LISA.
Unfortunately, no single event has been detected so far by the various PTA collaborations
and the relevance of this effect in a stochastic background has not been properly studied yet.

In this work, we adapt this formalism to the geometry of the LISA mission. Although the LISA arm length 
is significantly
shorter than typical pulsar distances, the high precision of laser interferometry may offer a unique opportunity to test
these corrections of cosmological origin.
According to the previous discussion we will be able to see an effect from $k_{eff}$ provided that the corresponding
wavelength is comparable or smaller than
the arm's length $L$. In the case of PTA, a typical wavelength from very  massive extragalactic black holes mergers
is of the order of light-year making pulsar distances sensitive to that effect. In the case of LISA, the arm length is
2.5 millions of kilometers suggesting that the effect would be
relevant for frequencies around or greater than 100 mHz. We will see this clearly in the remaining sections.

LISA loses sensitivity above 1 Hz because once gravitational wave wavelengths become
comparable to the arm length, the interferometer response rapidly collapses. At high frequencies,
the GW signal averages out along the arm, producing some cancellation. Instrument noise also rises in
this band, but the dominant reason is geometric: LISA is simply too large to detect short‑wavelength GW efficiently.
However, we will see that the enhancement just described makes frequencies between 1 Hz and 10 Hz particularly interesting
and we expect the effect to be both visible and useful.

We will focus first on individual arms and later
study correlations between two arms. We will determine the optimal angle of incidence $\alpha_{optim}$ that maximizes the effect that
would be correlated with the
redshift. This would represent a completely independent measure of the redshift, totally independent of measuring the arriving
frequency and comparing that result
with a modelization of the source. The extension to the full interferometric detector response is beyond the scope of this
work and is left for future studies.


\section{The LISA Mission}

The Laser Interferometer Space Antenna (LISA) is a future space-based gravitational wave observatory led by the European Space Agency (ESA).
Unlike ground-based detectors such as LIGO and Virgo, which are limited by seismic noise at low frequencies, LISA will operate in space,
prepared to detect gravitational waves in the milli-Hertz band ($0.1$ mHz to $1$ Hz). This frequency range is rich in astrophysical
sources, including supermassive black hole binaries, extreme mass ratio inspirals (EMRIs), and galactic binaries \cite{LISAreview,AmaroSeoane2017}
and somehow  bridges the gap between LIGO/Virgo ($10$--$10^3$ Hz) and Pulsar Timing Arrays ( $10^{-9}$--$10^{-6}$ Hz)
\begin{table}[ht]
\centering
\small
\setlength{\tabcolsep}{2pt}
\begin{tabular}{llll}
\toprule
\textbf{Phenomena} & \textbf{Redshift/Distance} & \textbf{Freq. ($f$)} & \textbf{Strain ($h$)} \\
\hline
\textbf{MBHB} & $z \approx 1-20$, $\sim 100$ Gpc & $10^{-5}- 10^{-1}$ Hz & $10^{-20} - 10^{-17}$ \\

\textbf{EMRI} & $z \approx 2$, $\sim 16$ Gpc & $1 - 100$ mHz & $10^{-22} - 10^{-21}$ \\

\textbf{GUCB} & $z \approx 0$, $1-20$ kpc & $0.1 - 100$ mHz & $10^{-23} - 10^{ -21}$ \\

\textbf{SOBHB} & $z \approx 0.02$, $\sim 100$ Mpc & $10$ mHz $ - 1$ Hz & $10^{-24} - 10^{-22}$ \\

\textbf{SGWB} & $z \gg 1000$, Horizon & $10^{-5} - 10^{-1}$ Hz & $10^{-12} - 10^{-9}$ \\
\hline
\end{tabular}
\caption{LISA Observational Phenomena summary\cite{AmaroSeoane2017}. MBHB: massive black hole binary; EMRI: extreme mass-ratio inspiral;
GUCB: galactic ultra-compact binary; SOBHB: stellar origin black hole binary; SGWB: stochastic gravitational wave background.}
\label{tab:lisa_summary}
\end{table}

The mission consists of three spacecraft arranged in an equilateral triangle constellation. They will trail the
Earth in its heliocentric orbit at a distance of approximately 50 million kilometers. The center of the constellation
traces an Earth-like orbit, while the plane of the triangle is inclined $60^\circ$ with respect to the ecliptic.

Crucially for this work, the three spacecraft are separated by an arm length of $L \approx 2.5 \times 10^6$ km.
While this distance is vast compared to Earth-bound interferometers, it is significantly shorter than the galactic scales
characteristic of PTA. 
Each spacecraft houses free-falling test masses  that act as the reference for geodesic motion. Lasers are exchanged
between the spacecraft to continuously monitor the distance between these test masses.

A passing gravitational wave distorts the spacetime metric, changing the proper distance between the test masses.
This change manifests as a phase shift in the laser light received by the remote satellite. LISA is capable of detecting
strain amplitudes as small as $h \approx 10^{-21}$, which corresponds to a displacement resolution in the
picometer range ($10^{-12}$ m). Standard interferometry is insufficient; instead, a technique known as
Time Delay Interferometry (TDI) \cite{Tinto2021} is used to cancel laser phase noise. For the purpose of the theoretical
analysis in this work, we will model the measurement as a timing residual accumulated over the  path along the
arm of length $L$.  After multiplying by $c$ this translates into an optical path difference. The locking in phase of the 
TDI can be be maintained for timescales of $\Delta T=1000$ s and this will be our basic integration time for the signal.


\section{Gravitational waves in a $\Lambda$CDM universe}

The analysis in \cite{espriupuigdomenech,espriualfaro} considers perturbations on Friedmann-Lemaitre-Robertson-Walker background. The propagation of waves is governed by the linearized equation of motion for metric perturbations in a FLRW background
that, after inclusion of the remaining densities, is given by Eq. \ref{linearized}. 
This equation can be solved order by order as an expansion on the cosmological densities. At leading order is
given by Eq. \ref{leading} and the solution at
the next-to-leading and next-to-next-to-leading
order can be found in \cite{espriu2021}. It is important to reiterate that beyond leading order the contribution
from the cosmological constant,  matter density and radiation density contribute in varying proportions so that the solution
is not expressable in terms of $H_0$ alone\cite{espriu2021}. However, this is probably beyond the
precision that can be attained at LISA.

It is also important to note that this expansion and the resulting solution are valid not only well inside the cosmological horizon,
but one has to make sure that corrections to the lowest order (i.e. $O(H_0)$) are under control. 
 In \cite{espriu2021} the modifications induced by the next and next-to-leading terms were examined and it was concluded that the approximation was
reasonably good for redshift $z<2$, approximately corresponding to 5 Gpc.

The physical observable sensitive to these corrections is the timing residual, $\tau_{GW}$. It represents the difference in
the arrival time of a photon (or electromagnetic pulse) due to the metric perturbation caused by the gravitational wave.
For a photon traveling from a source (or emitter) to an observer (receiver) separated by a distance $L$, the timing residual is given by\cite{MaggioreVol1}
\begin{equation}
\tau_{GW}(T) = -\frac{1}{2} \hat{n}^i \hat{n}^j \mathcal{H}_{ij}(T),
\label{eq:tau_def}
\end{equation}
where $\hat{n}$ is the unit vector along the path of propagation and $\mathcal{H}_{ij}$ involves the integral of the metric
perturbation $h_{ij}$ along the line of sight.

The timing residual is given by the integral
\begin{equation}
\mathcal{H}_{ij}(T) = \frac{L}{c} \int_{-1}^{0} h_{ij}^{FLRW} \left(T + \frac{xL}{c}, \vec{R}(x) \right) dx,
\label{eq:integral_general}
\end{equation}
where $x$ is a parametrization of the path, $T$ is the time at the receiver, and $\vec{R}(x)$ describes the spatial trajectory of
the photon. From now on we will set $c=1$.


\section{Optimization for the LISA configuration}

In this section, we adapt the general formalism to the specific geometry of the LISA mission. Our goal is to find the configuration
that maximizes the timing residual $\tau_{GW}$, thereby optimizing the potential for detecting the dependence on the cosmological parameters.
We consider the configuration depicted in the figure being $A$ and $B$ two of the LISA salellites and  $L$ being the arm length.
We complete the triangle with the source of the GW. Let $Z_A$ be the distance from the source to satellite $A$, and $\alpha$ be the
angle subtended at vertex $A$ with respect to the source.

\vspace{-0.2cm}

\begin{figure}[h!]
\centering
\includegraphics[scale=0.5]{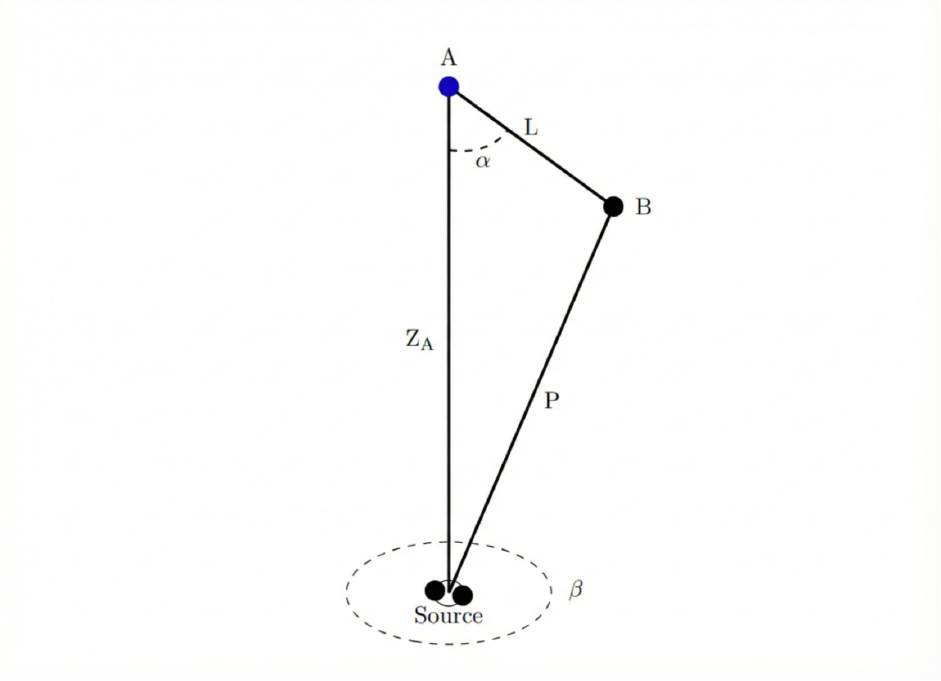}
\caption{Geometry of the system for LISA. The line AB represents one of LISA's arms.}
\label{fig:sample}
\end{figure}

The position along the LISA arm can be parameterized by a vector $\vec{R}(x)$, where the parameter $x$ runs from $-1$ to $0$.
The position vector is given by
\begin{equation}
\vec{R}(x) = \vec{P} + L(1+x)\hat{n},
\label{eq:position_vector}
\end{equation}
where $\vec{P}$ is the position of satellite $B$, and $\hat{n}$ is the unit vector along the arm direction. 

Given the astrophysical scales involved, we assume the condition $Z_A \gg L$. Under this approximation, the magnitude of the position
vector simplifies to
\begin{equation}
R(x) \approx Z_A + xL \cos(\alpha).
\label{eq:R_approx}
\end{equation}
On the other hand,
\begin{equation}
  T(x)= T_A +xL.
  \end{equation}
This approximation is sufficient to capture the leading-order effects of the cosmological corrections over the baseline of the detector.

The gravitational wave timing residual $\tau_{GW}$ is obtained by integrating the metric perturbation along the photon path between the satellites.
Substituting the FLRW metric perturbation into the timing residual definition and working with natural units ($c=1$),
we obtain the explicit expression for the timing residual:
\begin{equation}
\tau_{GW} = -\frac{L\epsilon}{2}\sin^{2}(\alpha) \int_{-1}^{0} \frac{1+H_0 T(x)}{R(x)}  \cos[\Theta(x)] dx,
\label{eq:tau_explicit}
\end{equation}
where $\epsilon$ represents the magnitude of the gravitational wave amplitude at the source (and we assume for simplicity that the non-zero
components of the polarization tensor are averaged and included in $\epsilon$). In short, $\epsilon/R(x)$ can be identified as
the strain at point $x$ along the AB line; $\alpha$ is the polar angle; $R(x)$ describes the photon path and $\Theta(x)$
is the phase of the wave along
the optical path. 

The phase $\Theta(x)$ is crucial for our analysis. It takes the form of a quadratic function:
\begin{equation}
\Theta(x) = A x^2 + B x + C,
\label{eq:phase_quadratic}
\end{equation}
where the coefficients $A$, $B$, and $C$ depend on the cosmological parameters $H_0$ and the geometry ($L, Z_A, \alpha$).
Specifically, the coefficients are found to be:
\begin{align}
A &= \frac{\omega H_0 L^2}{2} (\cos\alpha - 2) \cos\alpha, \\
B &= L\omega \left[1 -H_0 Z_A + (-H_0 T_A + H_0 Z_A - 1) \cos\alpha \right], \\
C &= \frac{\omega}{2} (-2 H_0 T_A Z_A + H_0 Z_A^2 + 2 T_A - 2 Z_A + \phi_0).
\end{align}
The phase $\phi_0$ has been introduced as initial condition to guarantee that at $x=0$ (i.e. the receiving
end of the AB line) the signal is zero until the gravitational wave arrives at time $T_A=Z_A$. Notice that this is absolutely
necessary because otherwise the constant $C$ is enormous and totally unphysical. It is worth noting that the constant
$A$ is totally negligible because is many orders of magnitude smaller than $B$, except when $B$ is zero, but the cosinus
is anyway extremely close to one in this case.

\subsection{Maximization via stationary phase}

The integral in question cannot be computed exactly -- it is related to Fresnel special functions-- but it can be determined
analytically with very good accuracy. We
can employ the stationary phase approximation \cite{ArfkenWeber}. The integral in Eq. \ref{eq:tau_explicit} is
highly oscillatory and the main contribution comes from the region where the phase is stationary.
The stationary point $x_0$ is found where the derivative of the phase vanishes, which corresponds to the apex of
the parabola $Ax^2 + Bx + C$:
\begin{equation}
x_0 = -\frac{B}{2A}.
\end{equation}
Evaluating the integral using the stationary phase method yields a result proportional to the cosine of the phase at the stationary point
\begin{equation}
\tau_{GW} \propto \cos\left( \frac{B^2}{4A} \right).
\end{equation}
To maximize this timing residual, we look for the condition where the argument of the cosine vanishes (or is minimized), which implies $B=0$.
Since $Z_A \approx T_A$ (since the travel time is dominated by the distance to the source), we arrive at the condition
\begin{equation}
\alpha_{optim} = 2 \arcsin \left( \sqrt{\frac{H_0 Z_A}{2}} \right).
\label{eq:alpha_optim}
\end{equation}
Thus observing the angle that maximizes the signal is tantamount to determining $H_0Z_A$.

When comparing the numerical and the stationary phase evaluations of the integral Eq. \ref{eq:tau_def},
it was seen in the case of PTA that the approximation was excellent\cite{espriupuigdomenech}. It was also seen that
the value of $\alpha_{optim}$ was remarkably independent of the frequency of the gravitational wave that means independence of
the nature of the source. This makes this correlation particularly interesting in our view.

As we will see later, in the case of LISA the stationary phase approximation describes the results accurately for frequencies above 100 mHz, but
the enhancement implied by  Eq. \ref{leading} is much less
marked than for PTA and this can be understood because around that frequency
$k_{eff}$ becomes commensurate with $1/L$, a circumstance not present in PTA. However, for frequencies around or above 1 Hz the correlation between
$\alpha_{optim}$ and $H_0$ is perfect and it is also independent of the frequency of the wave. All this in perfect agreement
with the physics insight described above.

\begin{table}
\centering
\begin{tabular}{c c c} 
 \hline \hline
 \textbf{Distance} ($Z_A$) & \textbf{Angle} (Theory) \\ 
 \hline
 1.0 Gpc & $40^\circ$ \\
 0.5 Gpc & $28^\circ$ \\
 0.1 Gpc & $12^\circ$ \\
 \hline \hline
\end{tabular}
\caption{Theoretical determination of the value of $\alpha_{optim}$ via stationary phase for $H=H_0$.}
\label{tab:sim_angles}
\end{table}


\section{Numerical analysis: one arm}

We will consider two cases: single sources of GW and stochastic background. For our purpose of identifying the combination of
comoving distance and Hubble parameter $Z_A H_0$ (that for moderate distances coincides approximately with the redshift) by
detecting the angular dependence of the source as seen by one of the ends of the arm with respecto the line defined by the
arm itself, it is more convenient to use only one arm.

\begin{figure}
 \centering
 \includegraphics[scale=0.3]{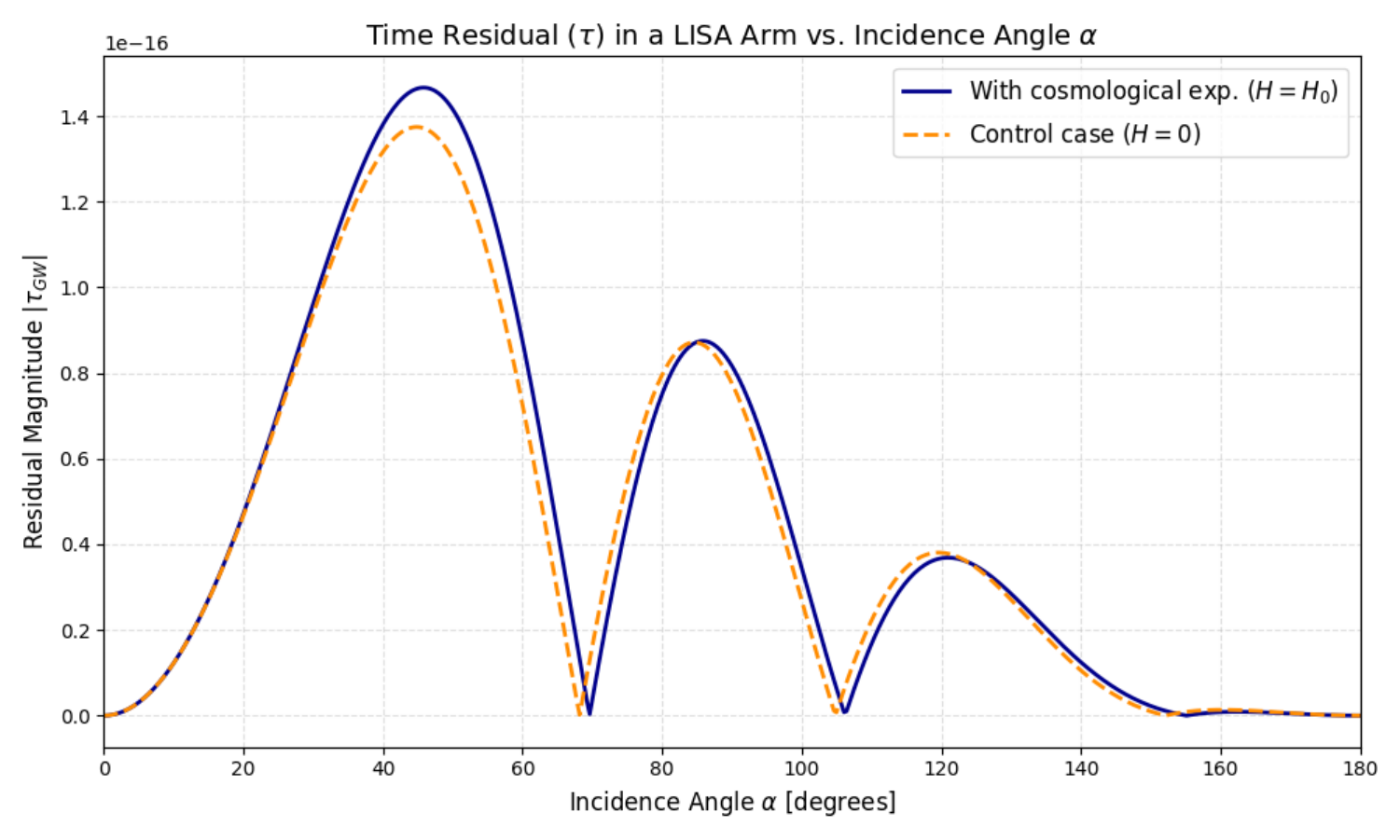}
 \includegraphics[scale=0.3]{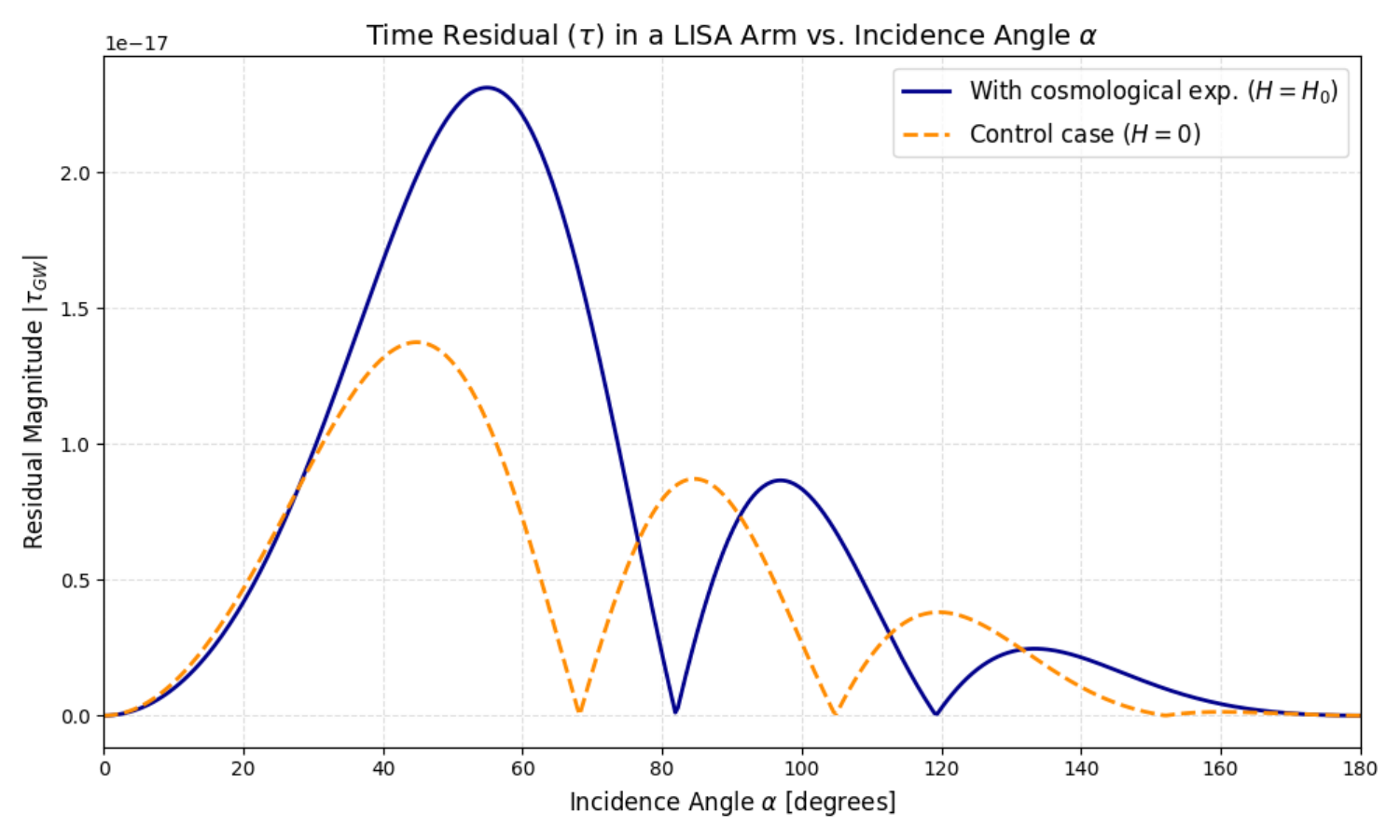}\\
 \includegraphics[scale=0.3]{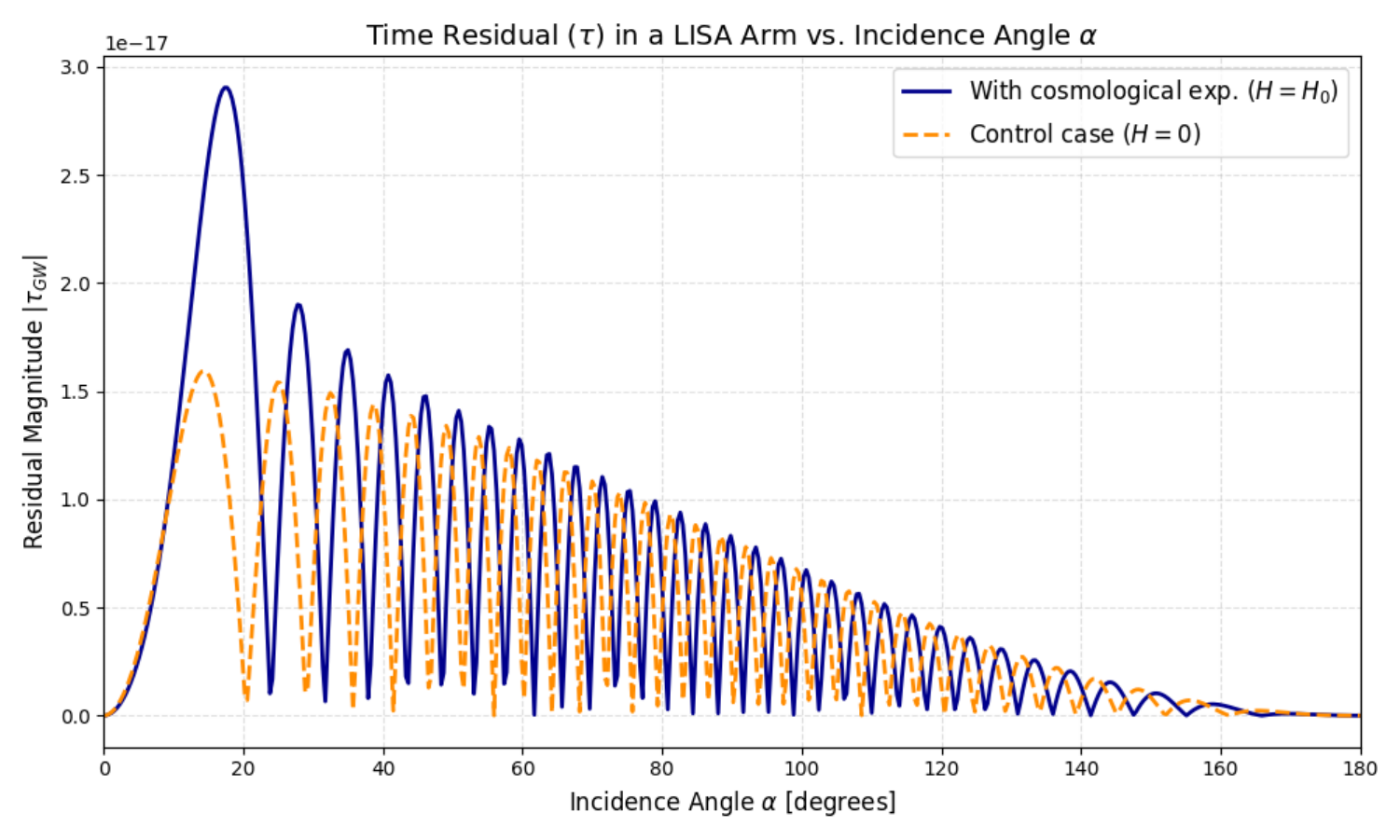}
 \includegraphics[scale=0.3]{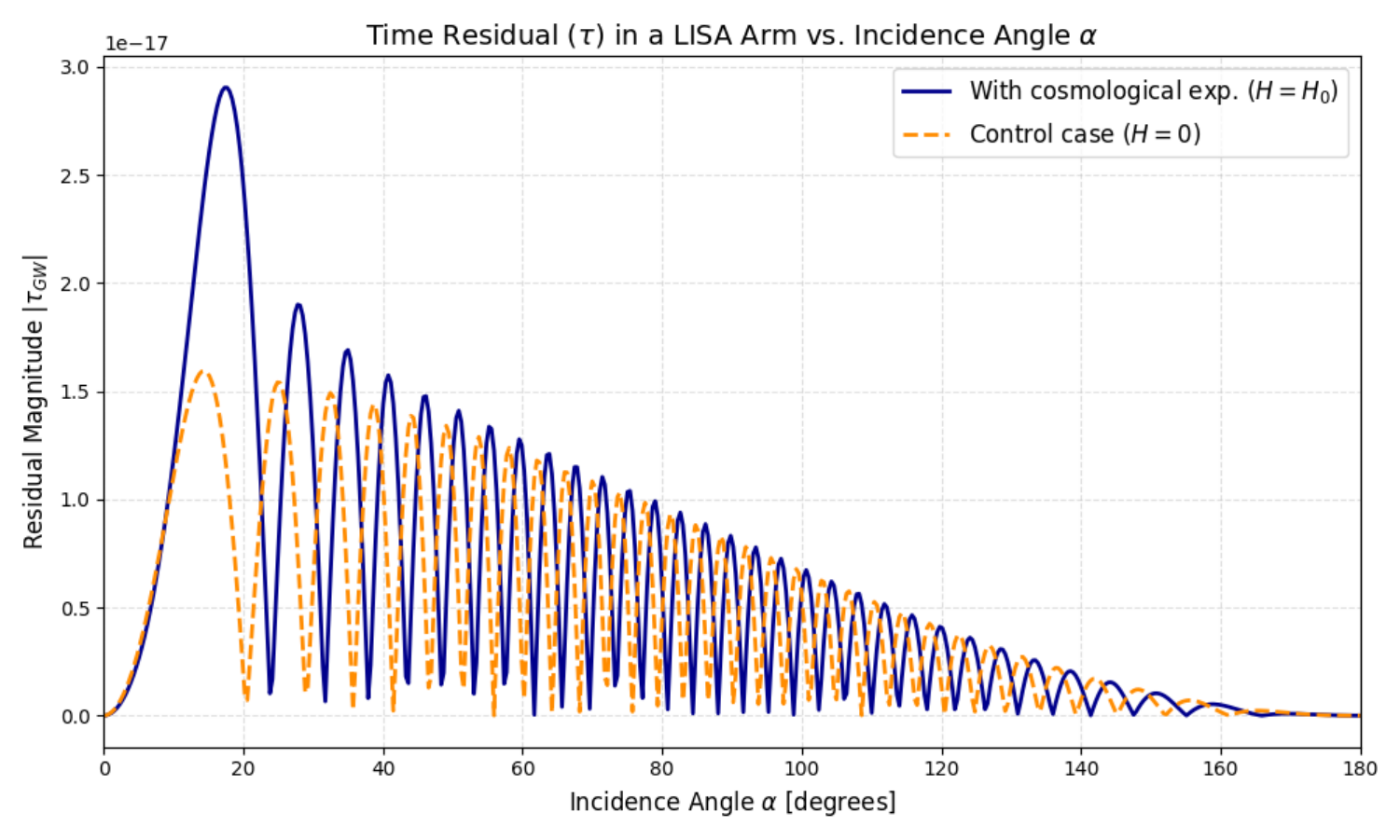}\\
 \caption{Comparison of the cases where $H=0$ (dashed line) and $H=H_0$ (solid line) for $Z_A$=0.1Gpc (left)
   and $Z_A$=1Gpc. The upper row corresponds to $\omega=0.6$ s$^{-1}$ and the lower one to $\omega= 6$ s$^{-1}$.
   In this plot and all the following ones, the value of the prefactor $\epsilon$ has been arbitrarily set to 1
 because it is only relative enhancements that will concern us.}
 \label{fig:wide_comparison}
\end{figure}

To begin with, we can compare the curves obtained for the time residual as a function of the angle subtended by the
source and the arm both for $H=0$ and $H=H_0$. The compromise value  $H_0 \approx 2.26 \times 10^{-18} \text{ s}^{-1}$ is used
throughout this study. In Figure \ref{fig:wide_comparison} we provide two snapshots for
two values of $Z_A$ and two frequencies.

For relatively close sources, the dependence on $H_0$ is very
mild for frequencies around or below 100mHz. It is more marked for more distant sources.  However, the situation changes
when we consider frequencies in the 1 Hz region. There is an enhancement both both distances, but still stronger
in the 1 Gpc case.

Of course it is possible to understand this by studying the integral leading to the timing residual, but the
physical reason is clear: at $f = 1$ Hz the fact that $k_{eff}^{-1} < L$ plays a role and the figures
show that the influence of the Hubble parameter goes beyond the modification of the redshift.

Let us now compare the enhancement at $\alpha_{optim}$ when the correct value for $k_{eff}$ is used to the
result obtained when one simple uses $k_{eff}= \omega_{eff}=\omega(1 - z)\simeq \omega/(1+z)$; that is, when only the
obvious redshift is taken into account. In this case the coefficients the new coefficients for $T=Z_A$
would be
\begin{equation}
A^\prime= \omega L^2 H_0 \cos\alpha (\cos\alpha -1) 
 \end{equation} 
\begin{equation}
B^\prime= \omega L (1-H_0Z_A)(1 -\cos\alpha)
\end{equation}

\begin{figure}
 \centering
 \includegraphics[scale=0.47]{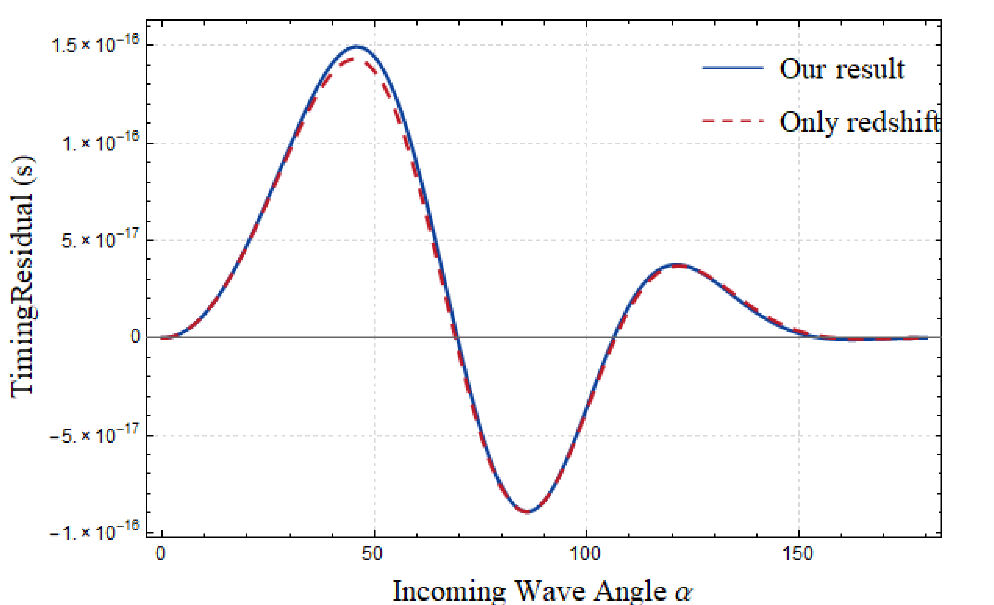}
 \includegraphics[scale=0.47]{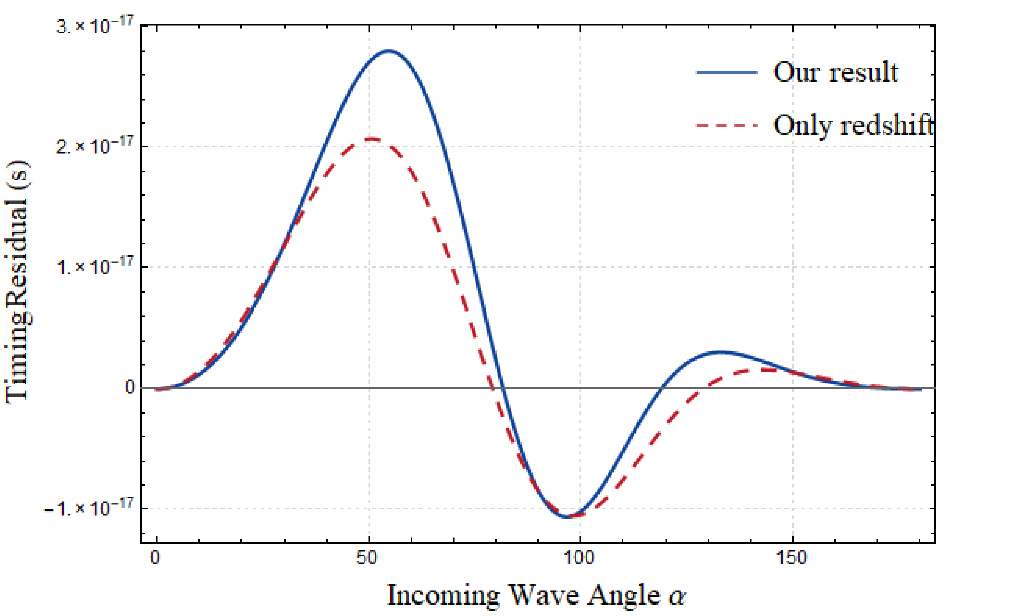}\\
 \includegraphics[scale=0.5]{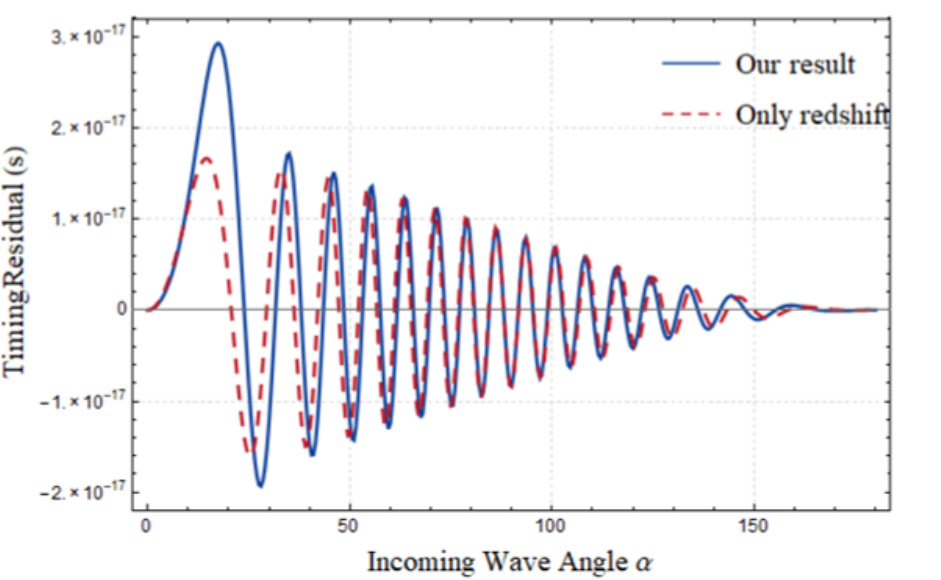}
 \includegraphics[scale=0.5]{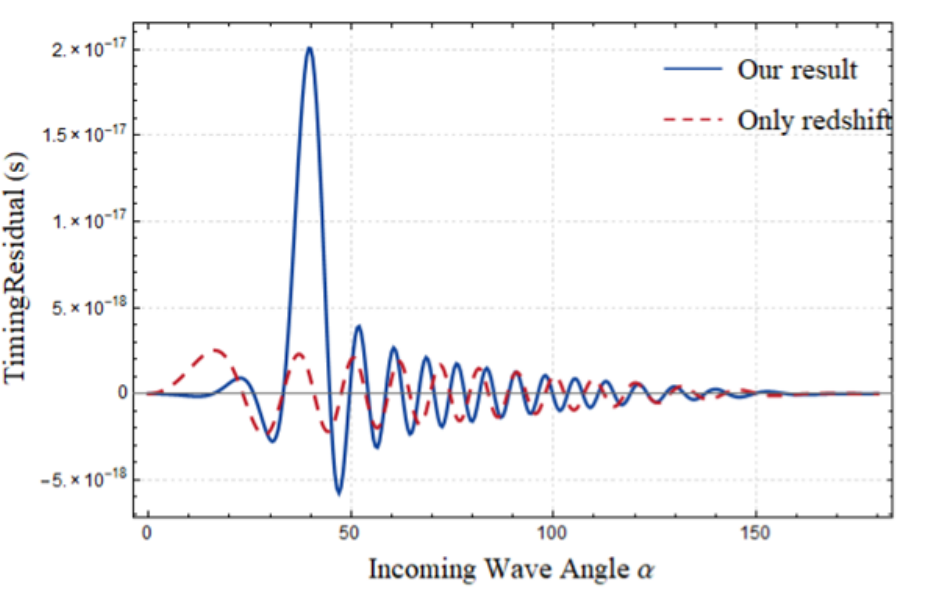}\\
 \caption{Comparison of the cases where the gravitational wave is described in terms of the redshift only (dashed line)
   and in terms of the proper solution of the wave equation to $O(H_0)$ (continuous line). The upper figures are for
   $f=100$ mHz, and the lower ones for $f=1$ Hz. The left(right)  figures correspond to 100 Mpc (1 Gpc).
   The difference is very noticeable for
   high frequencies and almost invisible around or below 100 mHz. As expected it becomes more marked for sources at large distances.
   Notice the expected signal degradation for $f=1$ Hz as discussed in the text that is compensated by the enhancement here described.}
 \label{fig:k=kversusk=w}
\end{figure}

The results are plotted in Figure \ref{fig:k=kversusk=w}.
It is worth noting that the stationary phase condition in the red-shift-only case also gives an enhancement but it is for
$\alpha=0$ or very close to zero, where the signal is geometrically supressed.

\begin{figure}
 \centering
 \includegraphics[scale=0.35]{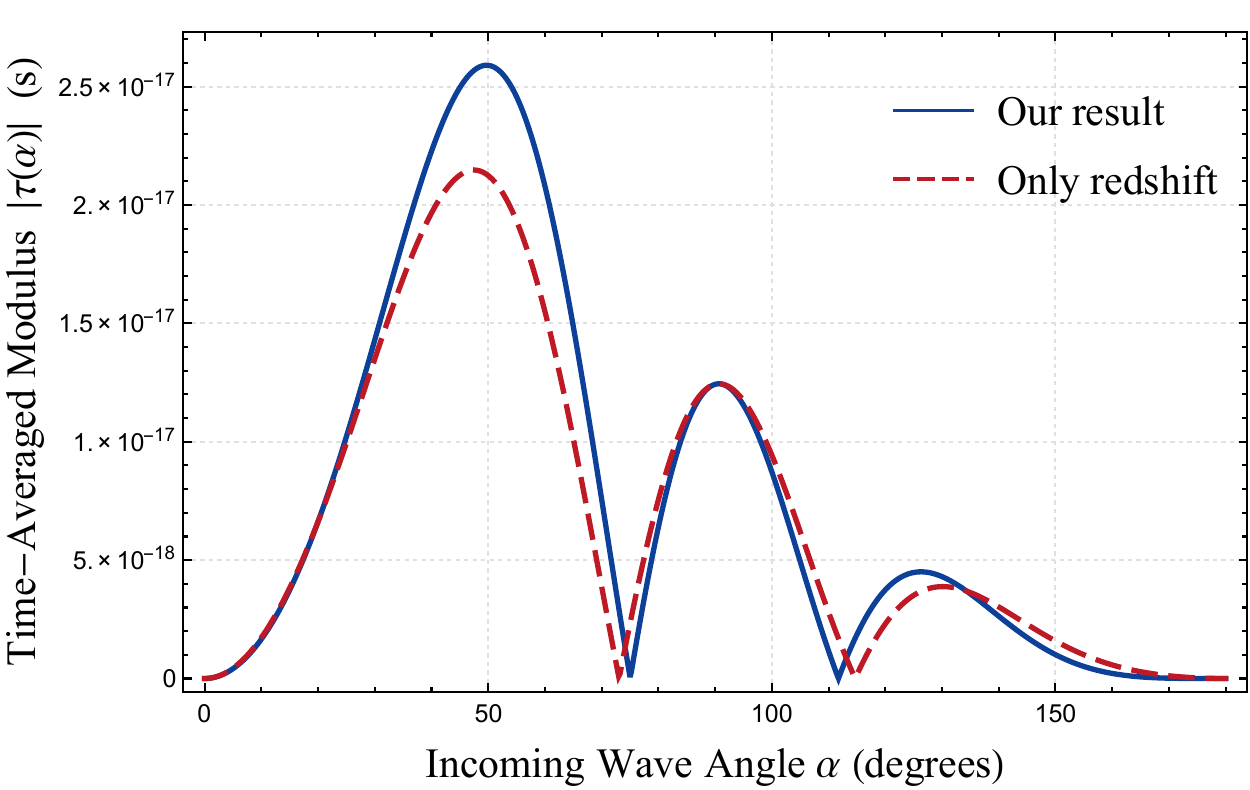}
 \includegraphics[scale=0.35]{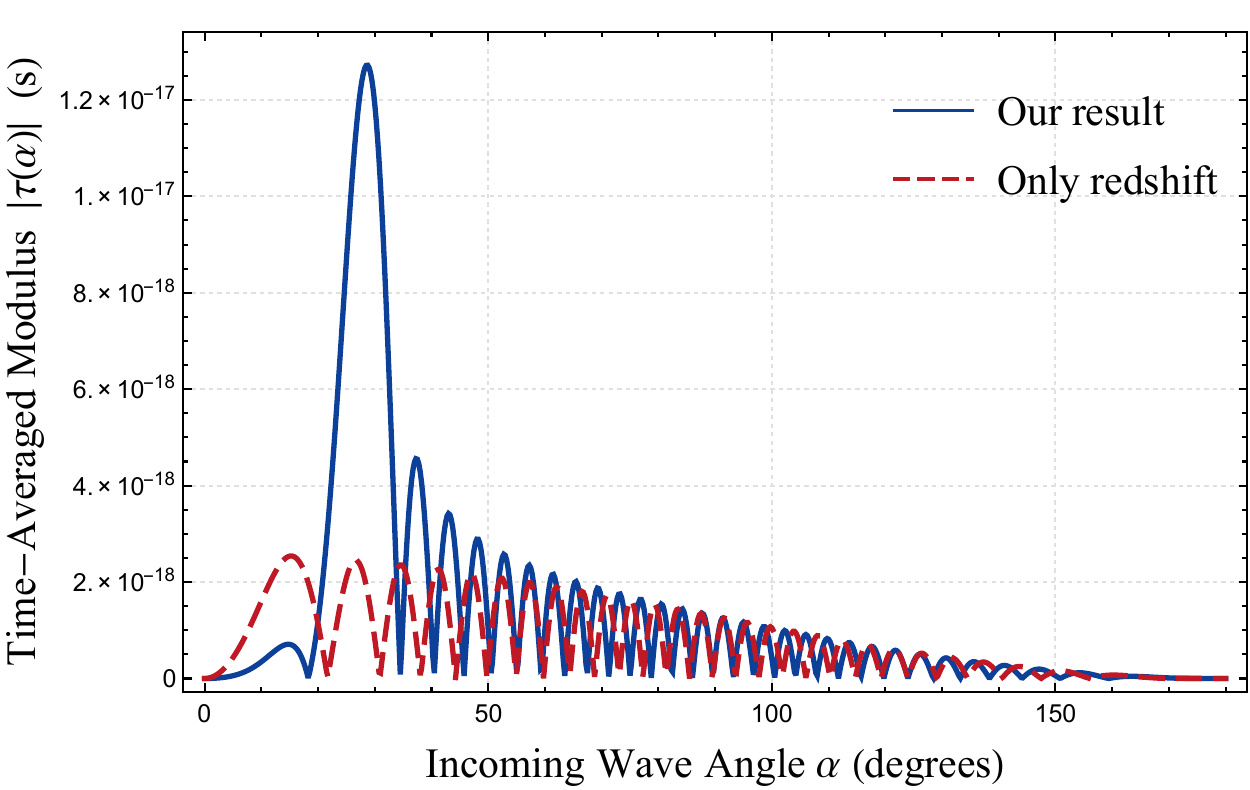}\\
 \caption{The time average results are not substantially different from those obtained from the snapshots at initial time $T=Z_A$. Here
   results for $Z_A= 0.5$ Gpc are presented using the standard compromise value $H_0=2.26 \times 10^{-18}$ s$^{-1}$.
   Left: $f=100$ mHz. Right: $f=1$ Hz. }
 \label{fig:timeaverage}
\end{figure}

The previous plots correspond to setting $T=Z_A$; in other words, we are studying the arrival of the signal. For long transients
we can also determine the average of the timing residual over a period of 1000 seconds, corresponding to the
laser phase locking. To do this we just have to replace everywhere $T_A=Z_A +t$ and average.
As an example, Figure \ref{fig:timeaverage} shows the result for $Z_A=0.5$ Gpc for two frequencies

\begin{table}
\centering
\begin{tabular}{c c c} 
 \hline \hline
 \textbf{Distance} ($Z_A$) & \textbf{Angle} ($\alpha_{optim}$) \\ 
 \hline
 5.0 Gpc & $98^\circ$ \\
 1.0 Gpc & $39^\circ$ \\
 0.5 Gpc & $28^\circ$ \\
 0.1 Gpc & $14^\circ$ \\
 \hline \hline
\end{tabular}
\caption{Determination of the value of $\alpha_{optim}$ for four comoving distances and a frequency $f=2$ Hz from an exact calculation.
  As already emphasized
  for frequencies in the Hz range and above the results are virtually independent of the frequency. The agreement with the theoretical
results is excellent.}
\label{tab:sim_angles2}
\end{table}

From these snapshots we conclude that even though the effect is by no means as dramatic as in PTA, the ``effective'' or observed
strain has a
very visible dependence on the relative angle subtended by the arm and the source; an effect that is of order one and therefore
measurable. If the $Z_A$ of the corresponding source is known, and taking into account the high sensitivity expected at LISA
it should be viable to determine $H_0$ with the precision of a few per cent. As it is well known, LISA will be able to determine
$Z_A$ with good precision once the chirp mass of the source is known, which in turn is determined from the relative variation
of the frequency $\dot f/f$.

We have observed that for frequencies in the Hz range and sources at about 1Gpc a variation of 10\% in the value of the
Hubble parameter translates into a variation of two
degrees in $\alpha_{optim}$. For sources at  5 Gpc the variation in $\alpha_{optim}$ grows to about 7 degrees. It becomes
less visible for sources at closer distances.
Given the expected accuracy of the LISA mission, this effect is surely measurable.

\section{Numerical analysis: two arms}

We will now move to a different way of determinining $H_0$ by the combined use of two arms (of course a combined use of the
three arms will increase the statistical significance, but we will not discuss it here).
We need first of all an estimate of the number of events $N$ that LISA will be able to detect per year as a function
of the redshift $z$; i.e. the function
$dN/dz$.

Two main populations will be dominating the detectable events of interest to us: massive black hole binary mergers (MBHBs)and
extreme mass-ratio inspirals (EMRIs). For MBHBs, which are among LISA primary targets, the redshift distribution
exhibits a robust, generic shape: At low redshift ($z \lesssim 1$), the number of detected events rises due to the
increasing cosmological volume and improving merger rates.
A broad maximum occurs at intermediate redshift ($z \sim 1-3$), corresponding to the peak of galaxy formation and merger
activity. Most simulations find a peak around $z  \approx  0.8-2$ \cite{mbhb}. At high redshift ($z \gtrsim 3-5$), the
distribution decreases and  although LISA may still detect sources up to $z\approx  10$ this is of no interest to us because
our methodology is not directly applicable there. The total number of MBHB detections is relatively modest up to about 20 per year
but uncertainties may raise this number up to $\sim 100$.

EMRIs involve a stellar-mass compact object (black hole, neutron star, or white dwarf) spiraling into a massive black
hole with mass $10^5-10^7 M_\odot$, with mass ratios $ \sim 10^{-5}-10^{-7}$. These systems produce long-duration,
information-rich signals but are intrinsically weaker than MBHB mergers. Unlike MBHBs, EMRIs are  biased toward low redshif with
a distribution expected to peak at $ z \sim 0.3-0.7$ and falling rapidly beyond $z=1$
This is because the GW amplitude scales with the small secondary mass, compromising  EMRI detectability.
A key result in the literature is that EMRIs could dominate LISA catalog. Most estimates lie in the 10-100
range but optimistic scenario predictions extend to the 10-1000 region.

Other types of phenomena such as stellar-origin black hole binaries (SOBHBs) are predominantly local and of no interest
for our purposes.

\begin{figure}
 \centering
 \includegraphics[scale=0.45]{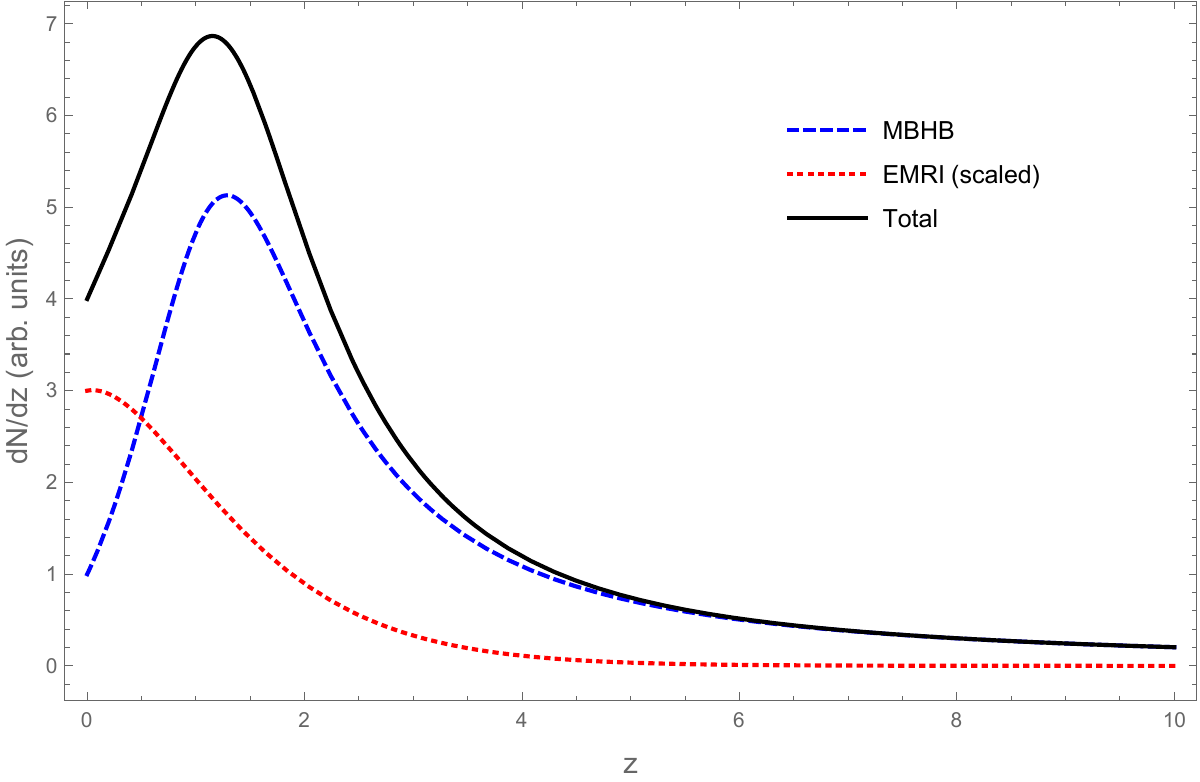} 
 \caption{Approximate redshift distribution of GW sources detectable by LISA. The total distribution is modeled as
   the sum of contributions from massive black hole binary mergers, peaking at $z \sim 1–2$, and extreme mass-ratio inspirals,
   concentrated at $z \lesssim 1$. The curves are qualitative  and based on typical results from the available
   the literature\cite{AmaroSeoane2017,klein,babak,mangliali}}
 \label{fig:numbersources}
\end{figure}

In order to demonstrate the method we will assume that LISA will be able to identify and determine the distance of some 500 sources
over its entire life span for each of the distance ranges that we have considered. Below we plot a comparison between the case $H=H_0$ and
$H=H_0$ as it would be detected by LISA after averaging 500 sources randomly distributed at three different comoving distances. The plots
show the sum of the signals measured by two LISA arms as a function of the respective angles subtended by the source and the two
arms. The frequency of the signal is assumed to be 1 Hz, recalling that above 100 mHz the enhancement is frequency independent.
\begin{figure}
 \centering
 \includegraphics[scale=0.7]{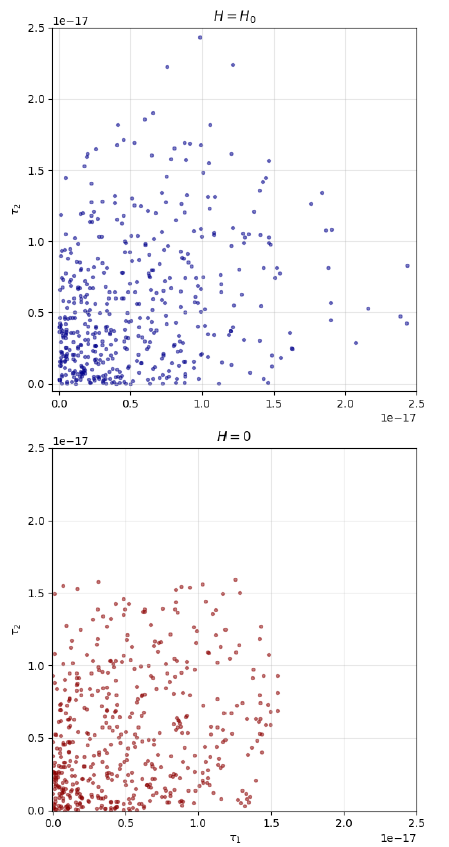}
 \includegraphics[scale=0.7]{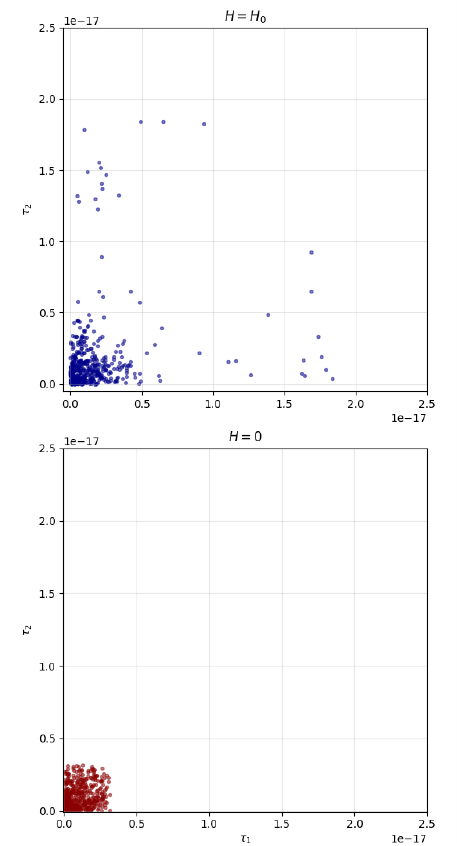}
 \includegraphics[scale=0.7]{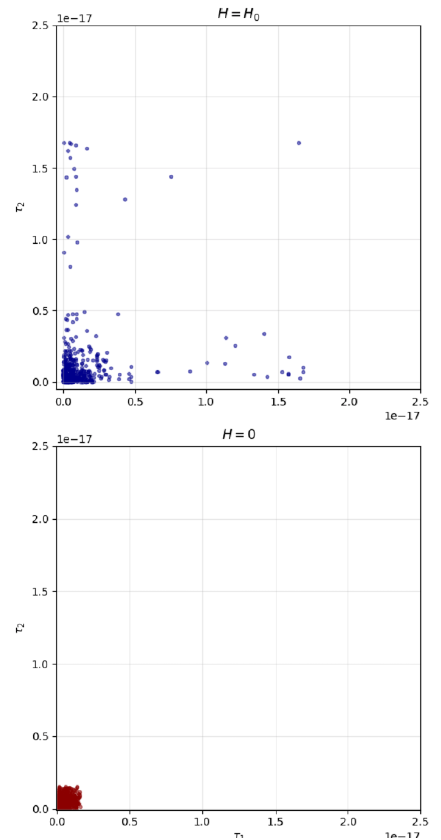}\\
 \caption{Comparison between the combined signal of two LISA arms for $H=H_0$ (upper figures) and $H=0$ (lower figures).
   The axes correspond to the timing residual as measured in the two arms and the plots describe a
   random distribution of 500 sources at three comoving distances: 100 Mpc, 500 Mpc and 1 Gpc (left to right).
   The frequency is taken to be is 1 Hz.
   It is interesting to see how a simple counting exercise is able to
   capture the fact that the universe is expanding and shows that the value of $H$ can be visualized.}
 \label{fig:averageangles}
\end{figure}

Next we will explore for a fixed frequency how the value of $H$ could be estimated for various distances. In the single arm case the
procedure consisted in profiling the intensity of the signal for various angles of the emitting sources and seeking for a maximum. This could be done
by observing similar sources located at various positions or, for long lived transient signals by taking advantage of the movement of the crafts.
Having three arms does of course improve the statistics.

Here our proposed method is simpler, we just register the sum of the time
delays of an observed source at a given moment and accumulate statistics in a two-dimensional plot with angles $\alpha_1$
and $\alpha_2$ describing the position of the source with respect to the two LISA arms; that is the angles
subtended by the source of the gravitational perturbation and the two arms.
In order to facilitate visualizing the signals, only
a fraction of the most intense signals is kept. The precise fraction is chosen dynamically to optimize the signal. In short, this
is a simple statistical method quite visual.

\begin{figure}\label{fig:exploringZ}
 \centering
 \includegraphics[scale=0.66]{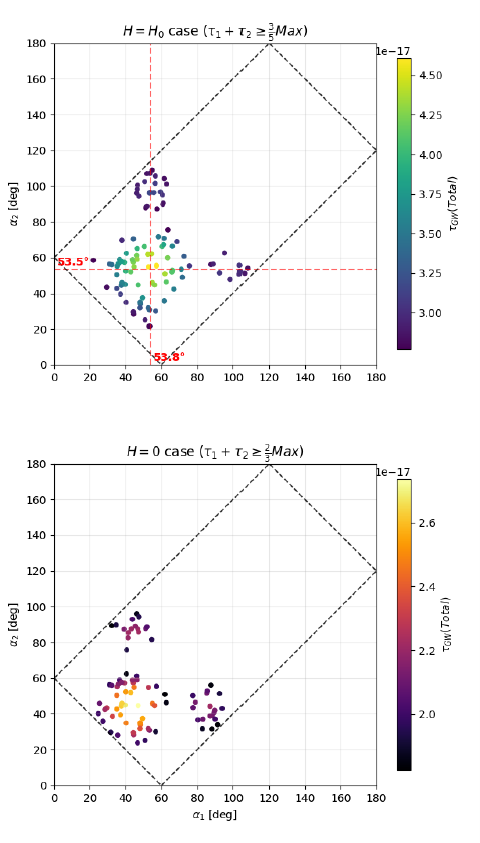}
 \includegraphics[scale=0.66]{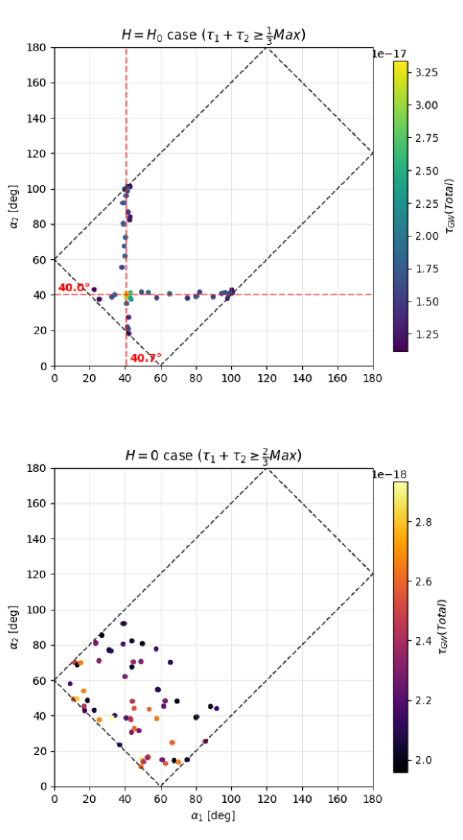}
 \includegraphics[scale=0.66]{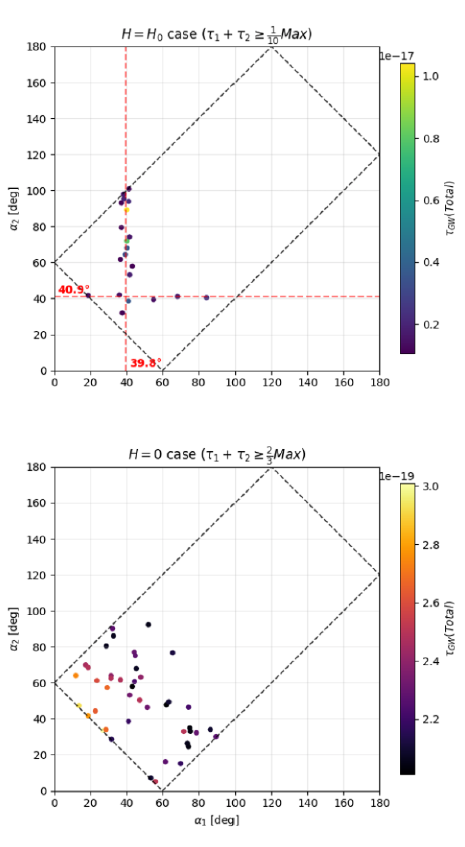}\\
   \caption{The plots depict the sum of the time residuals measured in two arms $|\tau_1| + |\tau_2|$ as a function of the respective angles
     $\alpha_1$ and $\alpha_2$ for a randomly distributed population of 500 GW sources.
     Only points in the inner rectangle are possible from a geometrical point of view. The signals are selected to
     fulfill minimum criteria (indicated in the graphic in each case) in order to make the plots understandable without needing to refer
     to the color coding (also included). The distance is always 1 Gpc and the frequency are left to right: 100 mHz, 1 Hz and 10Hz. The upper figures
   correspond to $H_0$ and the lower ones assume $H=0$}
\end{figure}

In the figures we observe that the situation changes drastically above the 100 mHz frequency (in agreement with the single arm results). The results
are virtually independent of $f$ and the difference between $H=H_0$ and $H=0$ is neatly visible. The cross-shaped axes in the latter case indicate
the fit to the signal, which is in full agreement with the analytical results of Tables II and III. For $f=100$ mHz no difference
is seen by the naked eye although the color coding signals some measurable differences.

For frequencies above the 100 m Hz range, the signals visibly show a cross-shape image that is visualized in the plots.
The position of this cross is an approximate determination of the value of $H_0$. Let us explore how the results depend on
the precise value of the Hubble parameter
\begin{figure}
 \centering
 \includegraphics[scale=0.28]{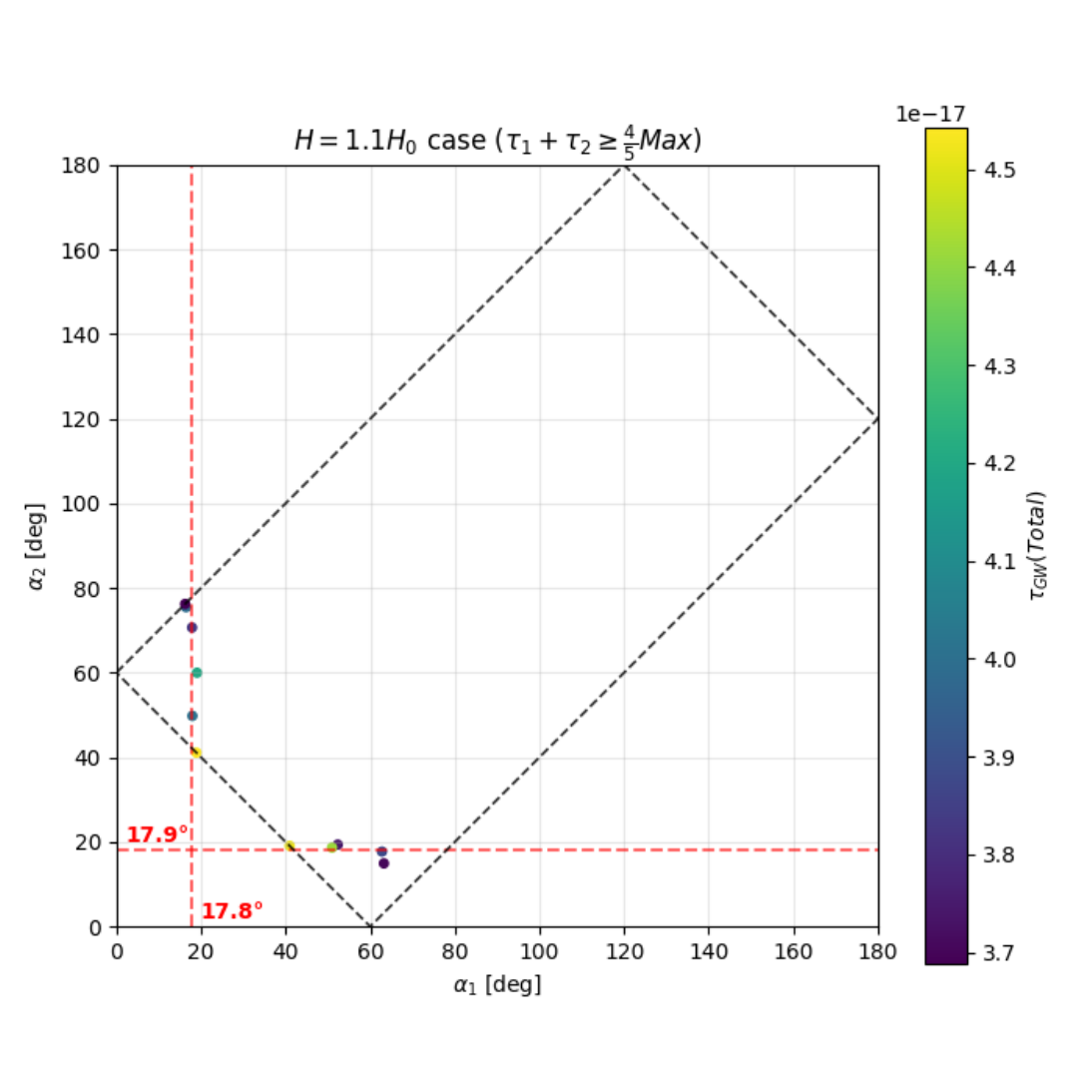}
 \includegraphics[scale=0.28]{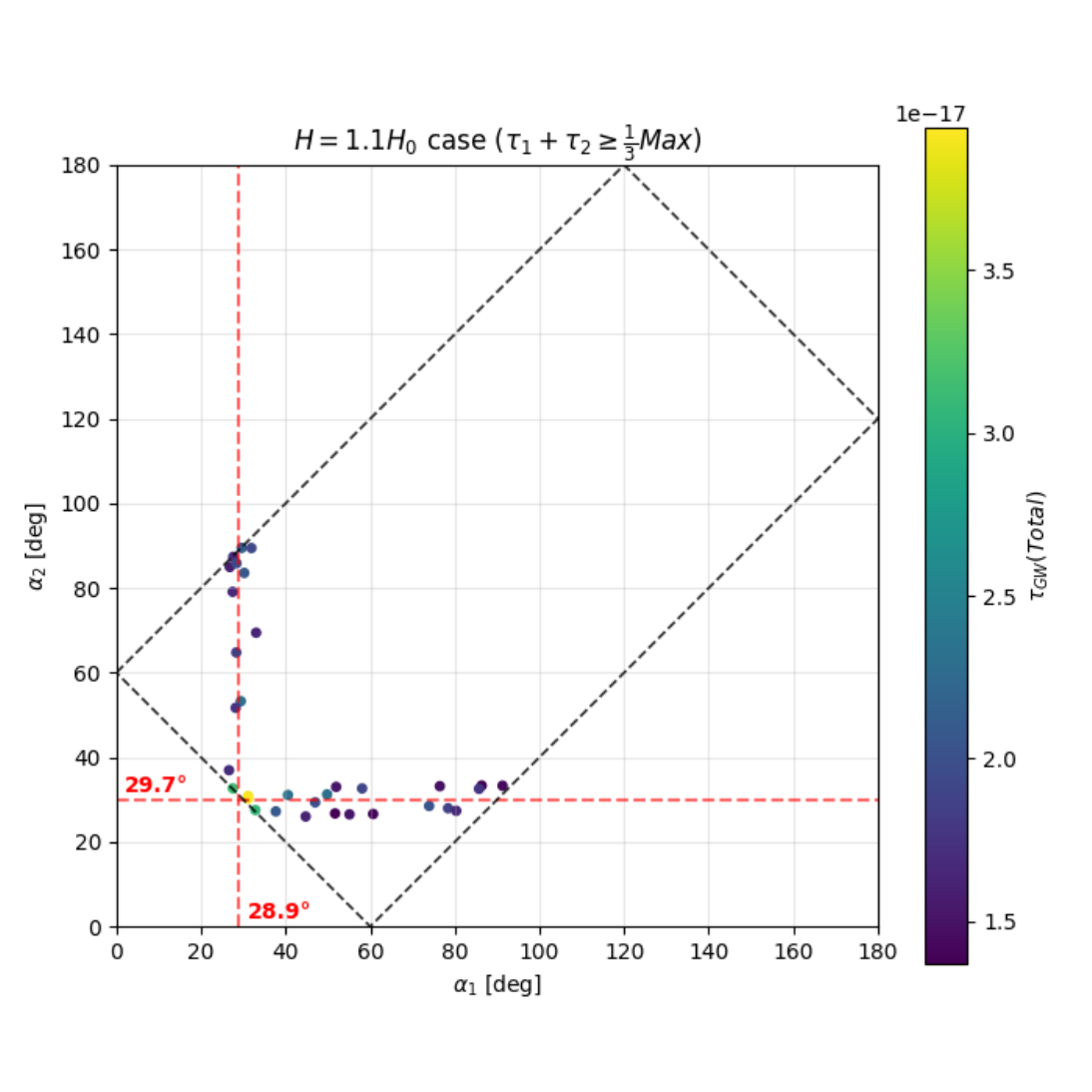}
 \includegraphics[scale=0.28]{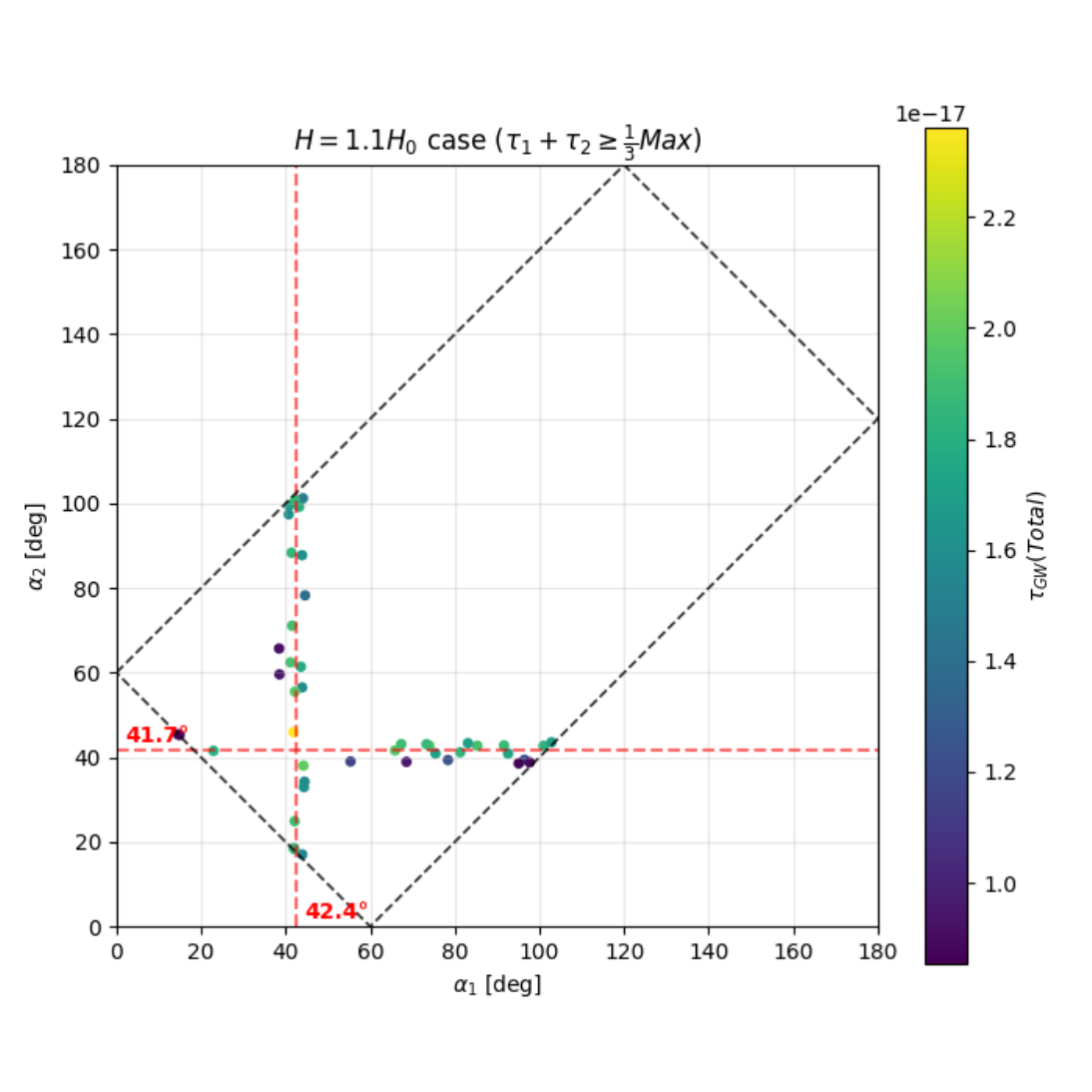}\\
 \includegraphics[scale=0.28]{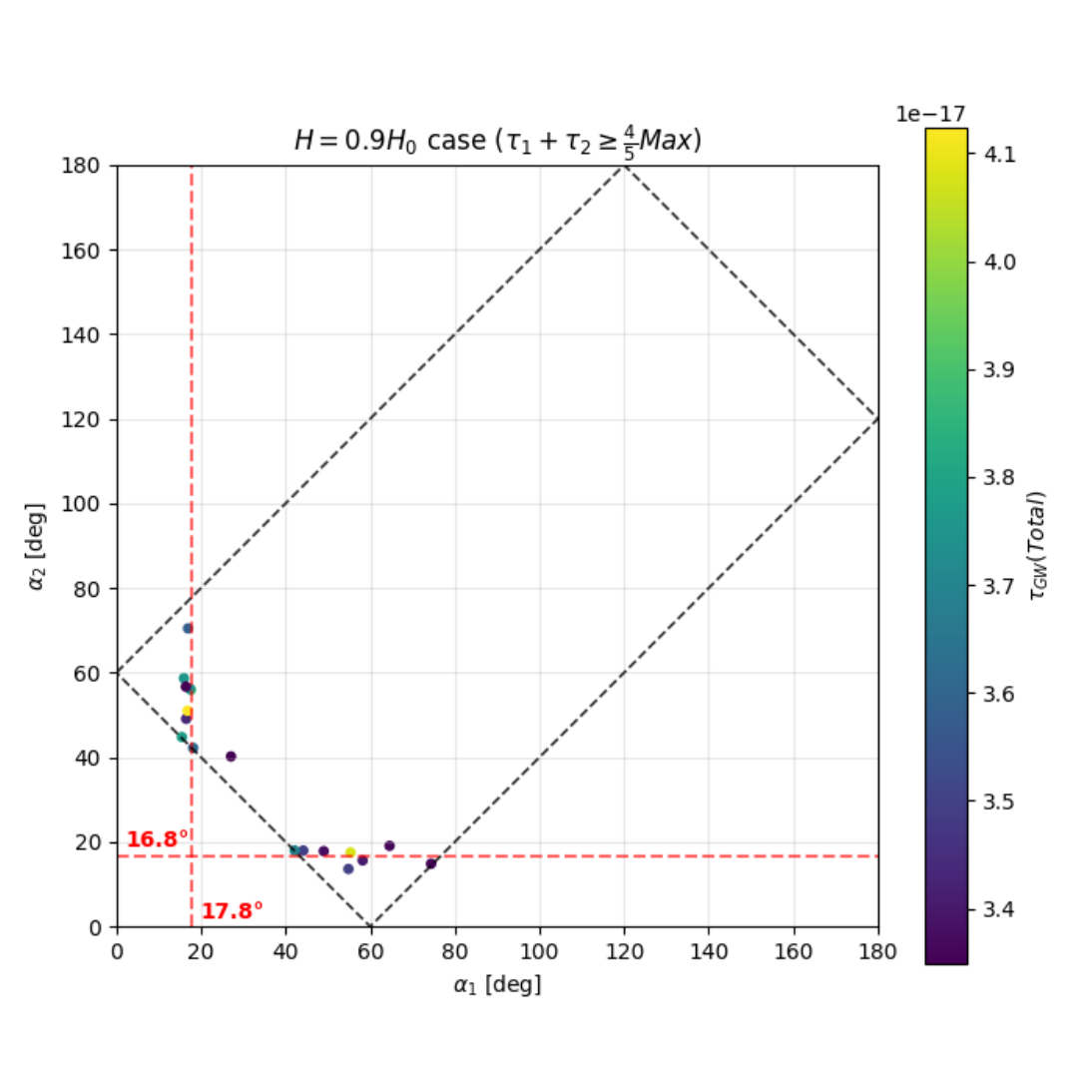}
 \includegraphics[scale=0.28]{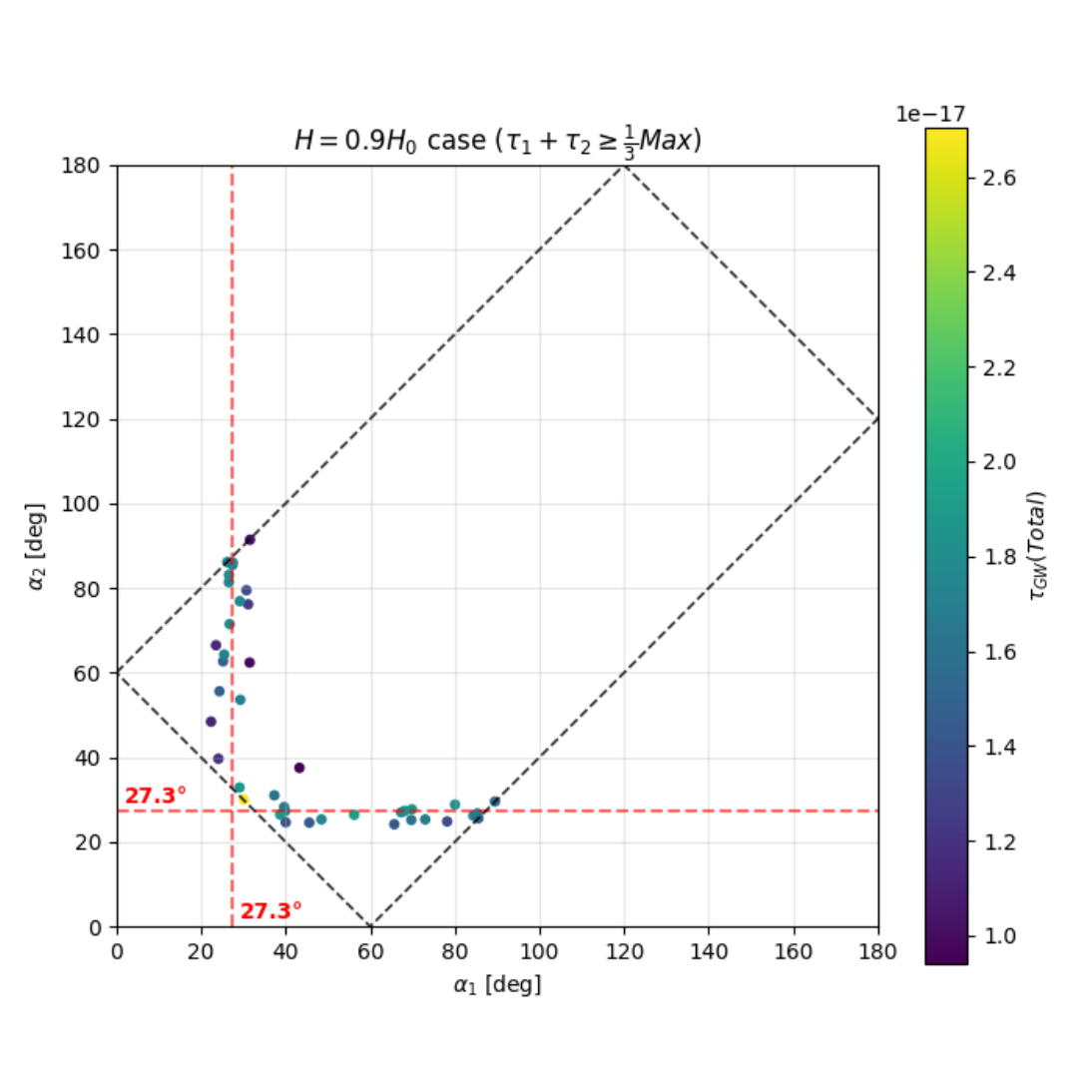}
 \includegraphics[scale=0.28]{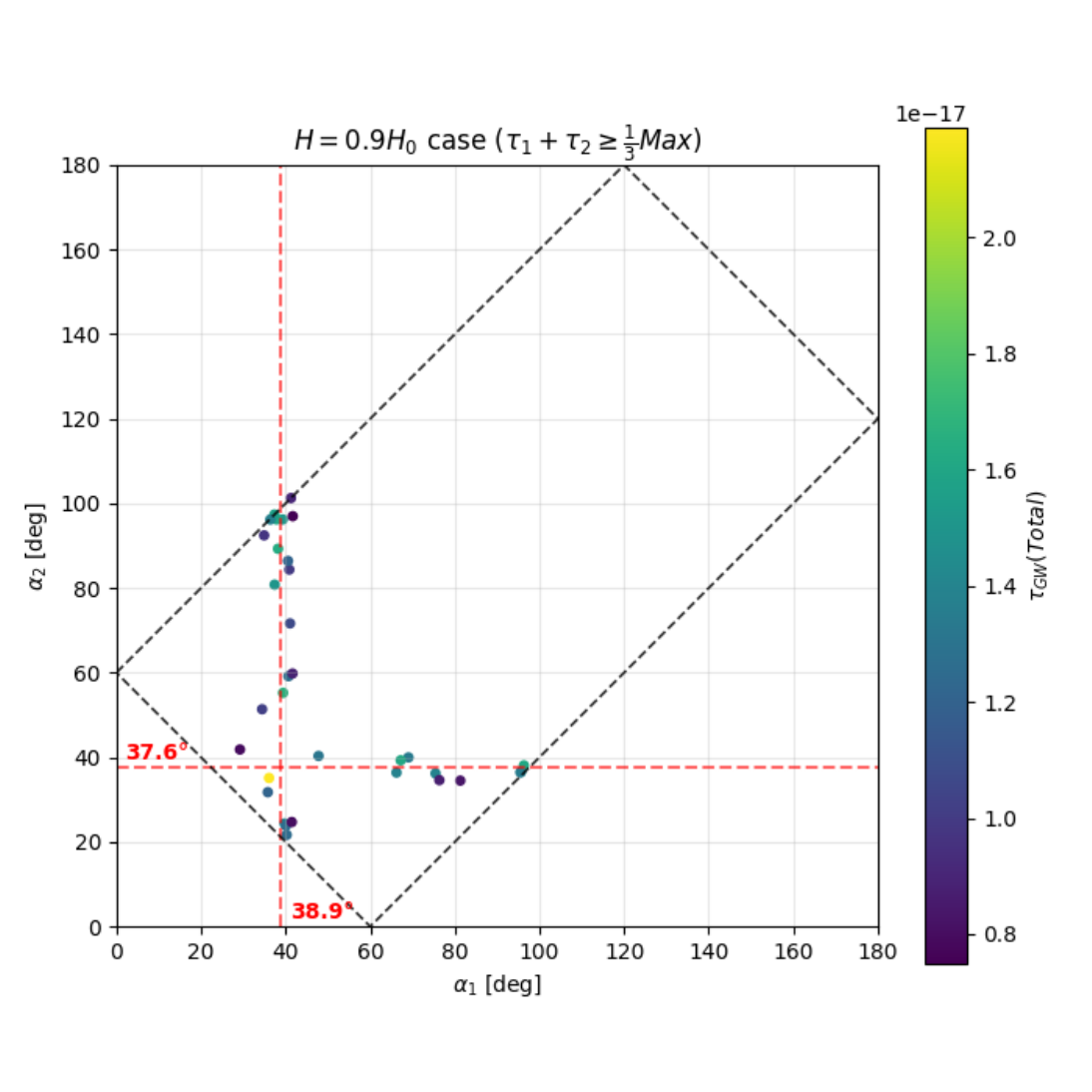}\\
 \caption{The sum of the timing delays measured by two LISA arms for several values of $H$ and comoving distances
   100 Mpc, 0.5 Gpc and 1 Gpc in a random background with 500 sources.
   Two different values of $H$ are plotted that differ in +10\% (upper) or -10\% (lower) from the conventional
   $H_0$. The frequency is 1 Hz. The results for $H_0$ are depicted in the previous figure.}
 \label{fig:exploringH}
\end{figure}
The figures show that the approximate determination of the Hubble parameter from this simple counting technique
will be more significant for comoving distances around 1 Gpc and above, where variations in $H$ reflect in
larger changes in $\alpha_{optim}$.

It should be mentioned that the above results all correspond to snapshots obtained at the arrival of the signal. Obviously
the time evolution, particularly for possible non-transient sources, will add to the statistics bearing in mind that the
orbital motion of the craft will continuously explore different angles for the same sources.
Needless to say that these results and figures need refinement before they can be turned into useful instruments for LISA operators.
They are provided only to show how the proposed method works and its potential capabilities.
We are well aware that choosing 500 random sources for each
comoving distance and frequency may not be realistic.

The most important caveat is that the angular modulation discussed in Secs. IV–V becomes very prominent only for GW frequencies
$f > 100 $ mHz, where the arm-length transfer function $ \mathcal{T}(f) \propto \sin(\pi f L)/(\pi f L)$ has not yet
fully suppressed the signal. However, LISA sensitivity in this band is expected to be degraded, and the astrophysical
sources radiating at these frequencies (principally stellar-origin black hole binaries) are predominantly at low
redshifts ($ z \lesssim 0.1 $), where the $ H_0 $-dependent correction is less obvious. A quantitative assessment of the
number of detectable sources is required to establish the final feasibility of the proposed method.
This remains an open question that surely requires more a detailed knowledge of the mission
than further theoretical work.

\section{Conclusions}

In this work, we have extended the formalism for cosmological corrections to gravitational wave propagation—previously developed for
Pulsar Timing Arrays—to the specific geometry of the LISA mission. Our analysis yields three primary findings, while also
highlighting the practical challenges that must be addressed for this method to reach its full potential.

First, we have shown that the phase shift induced by the Hubble parameter produces a distinct geometric signature.
Unlike standard redshift effects, which are degenerate with source properties, the $H_0$-dependent modulation of the timing residual
is maximized at an oblique incidence angle $\alpha_{\mathrm{optim}}$. Using the stationary phase approximation, we derived the scaling
relation $\alpha_{\mathrm{optim}} = 2\arcsin(\sqrt{H_0 Z_A / 2})$, which is remarkably independent of the gravitational wave frequency
for $f > 100$ mHz. Numerical integrations confirm this behavior, demonstrating that measuring this optimal angle provides
a direct handle on the combination $Z_A H_0$ without requiring a separate redshift determination. For single, well-localized
sources, this translates into a potential determination of the Hubble parameter.

Second, we have explicitly characterized the frequency-dependent nature of the signal. The enhancement becomes clearly visible
only above the 100 mHz threshold; below this frequency, the modulation is smeared out and the effect is largely degenerate
with geometric projection factors. This introduces a central tension: the angular modulation is most prominent precisely in the
band where LISA arm-length transfer function begins to suppress the response, and instrument noise may degrade sensitivity.
While our results suggest that the geometric enhancement may partially compensate for this suppression, we emphasize that a
rigorous signal-to-noise assessment—incorporating the full LISA sensitivity curve and the Time-Delay Interferometry
(TDI) response—is required to confirm the observability of individual events in this regime.

Third, we have examined the statistical prospects using populations of sources. For frequencies in the Hz range, the two-arm
correlation analysis produces a distinct cross-shaped pattern that is sensitive to the value of $H_0$, with variations of $\pm10\%$
leading to clearly distinguishable shifts in the aggregate timing residuals. However, this conclusion must be tempered by the
astrophysical realities of the LISA catalog. The sources that radiate most strongly above 100 mHz are predominantly
stellar-origin black hole binaries, expected to be largely local and  showing a weaker Hubble parameter correction.
Conversely, the massive black hole binaries and EMRIs that probe cosmological distances predominantly emit in the milli-Hertz band,
where our predicted geometric enhancement is less pronounced. This suggests that the optimal observational strategy may lie
in intermediate regimes or in stacking many lower-significance events, rather than relying solely on high-frequency bright sources.

Looking forward, several critical steps are needed to translate this proof-of-principle into a practical LISA measurement.
The most immediate is the incorporation of the full interferometric detector response, including TDI combinations and realistic
noise budgets. Additionally, a more detailed modeling of the source distribution, selection biases, and foreground confusion
(particularly from unresolved galactic binaries) will be essential to quantify the accessible parameter space.
Nonetheless, the method presented here offers a complementary route to measuring $H_0$ that is independent of the electromagnetic
counterpart or standard siren calibrations. If the technical and astrophysical challenges can be overcome, LISA may not only detect
the cosmological expansion but also constrain its present rate using an entirely new observational channel.

\section*{Acknowledgements}
 The financial support from
the State Agency for Research of the Spanish Ministry of Science, Innovation and Universities through the “Unit of Excellence Maria de Maeztu 2025-2028”
award to the Institute of Cosmos Sciences (CEX2024-001451-M) and through project PID2022-136224NB-C21 is acknowledged. We also acknowledge
the suppport of the Catalan Government through grant 2021-SGR-249.

\end{document}